\documentclass[aps,prl,twocolumn,superscriptaddress,10pt,floatfix]{revtex4-1}
  
\usepackage{graphicx}
\usepackage{hyperref}
\usepackage{booktabs}

\usepackage[utf8]{inputenc}

\renewcommand{\approx}{\sim}
\begin{document}

\title{Wet etch methods for InAs nanowire patterning and self-aligned electrical contacts}

\author{G.~F\"{u}l\"{o}p}
\affiliation{Department of Physics, Budapest University of Technology and Economics, and Condensed Matter Research Group of the Hungarian Academy of Sciences, Budafoki \'{u}t 8, 1111 Budapest, Hungary}

\author{S.~d'Hollosy}
\altaffiliation[present address: ]{Hightec MC AG,
Fabrikstrasse 9,
5600 Lenzburg, Switzerland}
\affiliation{Department of Physics, University of Basel, Klingelbergstrasse 82, CH-4056 Basel, Switzerland}
    
\author{L.~Hofstetter}
\altaffiliation[present address: ]{ABB Schweiz AG,
Corporate Research,
Segelhofstrasse 1K,
5405 Baden-Dättwil, Switzerland}
\affiliation{Department of Physics, University of Basel, Klingelbergstrasse 82, CH-4056 Basel, Switzerland}

\author{A.~Baumgartner}
\affiliation{Department of Physics, University of Basel, Klingelbergstrasse 82, CH-4056 Basel, Switzerland}
        
\author{J.~Nyg{\aa}rd}
\affiliation{Center for Quantum Devices \& Nano-Science Center, Niels Bohr Institute, University of Copenhagen, Universitetsparken 5, DK-2100
Copenhagen, Denmark}

\author{C.~Sch\"{o}nenberger}
\affiliation{Department of Physics, University of Basel, Klingelbergstrasse 82, CH-4056 Basel, Switzerland}

\author{S.~Csonka}
\email{csonka@mono.eik.bme.hu}
\affiliation{Department of Physics, Budapest University of Technology and Economics, and Condensed Matter Research Group of the Hungarian Academy of Sciences, Budafoki \'{u}t 8, 1111 Budapest, Hungary}

\begin{abstract}
Advanced synthesis of semiconductor nanowires (NWs)  enables their application in diverse fields, notably in chemical and electrical sensing, photovoltaics, or quantum electronic devices. In particular, Indium Arsenide (InAs) NWs are an ideal platform for quantum devices, e.g. they may host topological Majorana states. While the synthesis has been continously perfected, only few techniques were developed to tailor individual NWs after growth. Here we present three wet chemical etch methods for the post-growth morphological engineering of InAs NWs on the sub-100 nm scale. The first two methods allow the formation of self-aligned electrical contacts to etched NWs, while the third method results in conical shaped NW profiles ideal for creating smooth electrical potential gradients and shallow barriers. Low temperature experiments show that NWs with etched segments have stable transport characteristics and can serve as building blocks of quantum electronic devices. As an example we report the formation of a single electrically stable quantum dot between two etched NW segments.
\end{abstract}

\maketitle

\section{Introduction}

In the last two decades tremendous progress has been made in the synthesis of semiconductor nanowires (NWs). High quality crystals with a precise control of the geometry, composition and chemical properties enabled their application in various fields, like nanoelectronics, photonics, mechanical and biological systems, sensors or energy harvesting applications \cite{lieber2007functional,dasgupta201425th}.
III-V semiconductor NWs are particularly interesting for future quantum electonics applications: the strong electron confinement potential, the large spin-orbit interaction (SOI) in InAs and InSb and the possibility to create electrical contacts and strong superconducting proximity effect allowed the realization of exciting quantum devices. Examples are spin-orbit qubits, where the strongly coupled spin and orbital degrees of freedom of the electrons confined in quantum dots (QDs) represent quantum information \cite{nadj2010spin}, Cooper pair splitters, which generate spatially separated but spin entangled electron pairs \cite{hofstetter2009cooper,hofstetter2011finite,fulop2014local,das2012high}, or quantized charge pumping at GHz frequencies \cite{d2015gigahertz}. Recently intense activity was aimed at artificially creating topologically protected Majorana fermion states \cite{mourik2012signatures,alicea2010majorana,lutchyn2010majorana,oreg2010helical,das2012zero}. The latter relies on engineering a p-wave superconducting NW segment, exploiting SOI and the proximity effect, which is a key step towards topologically protected quantum computing. Ongoing experimental challenges call for exploring new fabrication strategies and the extension of the toolbox for NW-based circuit architectures.

Usually, the removal of the native semiconductor oxide on the surface of the NWs requires an etch step. This step is often also used to create a strong electronic doping of the semiconductor, to avoid Shottky barriers and obtain large carrier densities. However, wet chemical etch can also be used for the shaping of more sophisticated structures. For example wet etching has been used to create nanogaps in InAs--InP \cite{schukfeh2014formation} and InAs--GaAs \cite{kallesoe2010integration} axial heterostructure NWs, taking advantage of the material selectivity of the etch process, and in composite GaAs~2DEG--InAs NW structures for quantum dot and quantum point contact (QPC) formation \cite{shorubalko2008nw-qpc}. Furthermore, wet etching was used for the partial removal of the Al shell on InAs--Al core-shell NWs \cite{ziino2013epitaxial,krogstrup2015epitaxy}.

In this paper we present three wet chemical etching techniques for the post-growth patterning of homogeneous InAs NWs, two of which provide direct ways to realize self-aligned electrical contacts, while the third method results in a shallow gradient in the NW geometric profile as well as in the electrical potential. In addition to the detailed description of the fabrication methods, we also report the transport characteristics of the created structures. These methods allow for example the formation of smooth adiabatically changing constrictions in the NW or a narrowing of NW segments next to metallic contacts, highly relevant for the engineering of the electronic structure in nanometer scale quantum devices. 

\begin{figure}
\includegraphics{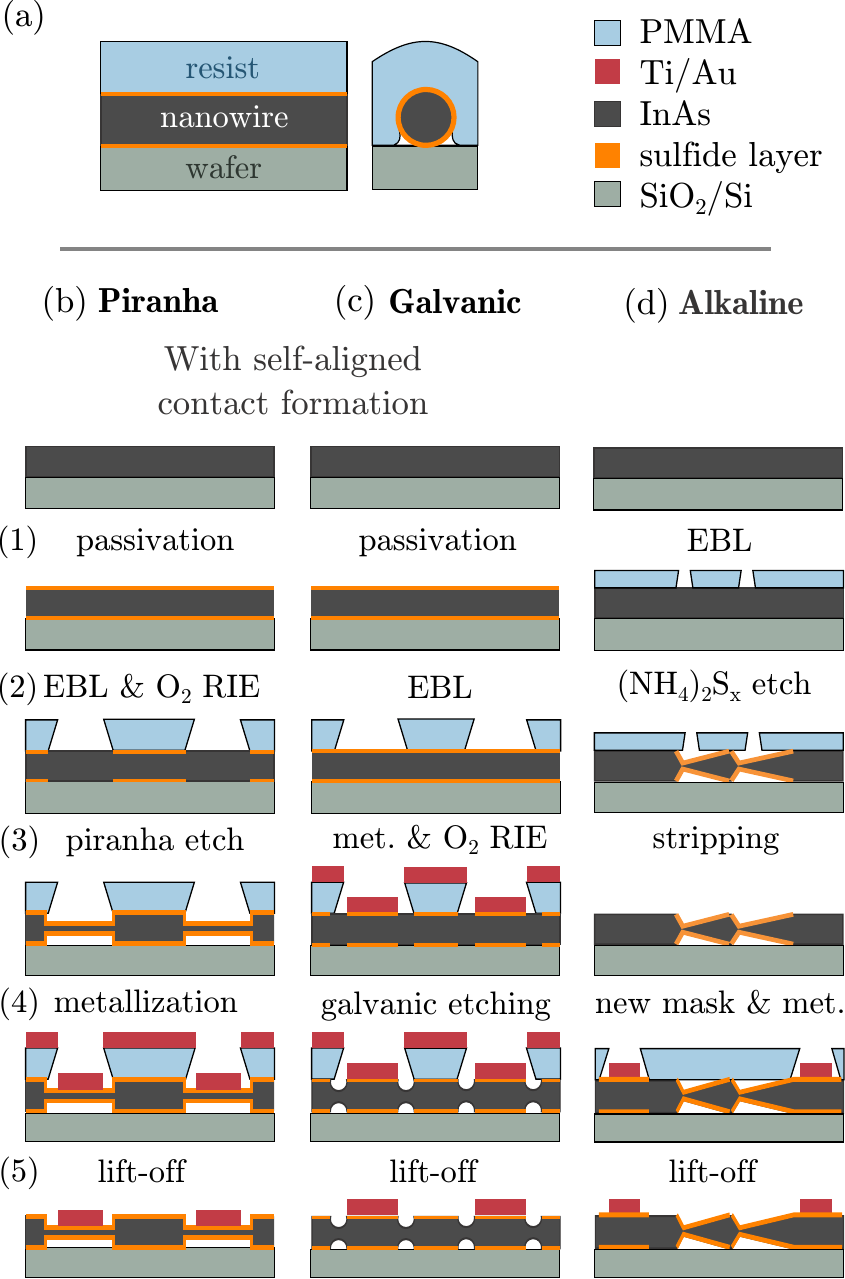}

\caption{Schematic, not to scale illustration of the etching methods. (a) A NW laying on the silicon substrate, covered by a resist layer, cross sections taken parallel and perpendicular to the NW axis. Due to the imperfect coverage of the resist the NW is not protected close to the NW--wafer boundary against liquids penetrating in the mask openings. (b-d) Fabrication procedures yielding a patterned NW with electrical contacts, using  (b) piranha, (c) galvanic and (d) alkaline etching.}
\label{fig1}
\end{figure}

\section{Experiments}

We have investigated the NW samples fabricated using all three methods by a scanning electron microscope (SEM). In case of the galvanic and alkaline methods, we have also carried out low-temperature transport measurements on the NW devices equipped with electrical contacts. First we present the generic elements of the sample fabrication and the overview of the three wet etch methods. The fabrication processes and the experimental results of each method are discussed in detail in the subsequent sections. 

\subsection{NW deposition and electron beam lithography}

InAs NWs were grown using molecular beam epitaxy \cite{madsen2013experimental,shtrikman2009method}  and dispersed in isopropyl alcohol.  A small amount of the dispersion was dropped onto an oxidized silicon substrate using a micropipette,  the solvent drying up left the NWs randomly deposited on the surface. The  NWs were localized using metallic markers fabricated beforehand and a resist mask was created by electron beam lithography (EBL).

In all methods polymethyl methacrylate (PMMA) was used as an e-beam resist. In case of the piranha and galvanic methods the resist thickness after spin coating and baking was $\sim330$~nm, while in the alkaline method, for the etching step the PMMA thickness was reduced to $\sim150$~nm which allowed us to achieve narrower etch windows. For the same reason we used an acceleration voltage of 30~kV for the electron exposure in the latter case,  20~kV in all others. 

\subsection{Overview of the wet chemical etch methods}
After their growth the NWs are transferred onto a silicon wafer and exposed to one of the three processes illustrated in Figure~\ref{fig1}. Two of these methods, shown in Figure~\ref{fig1}(b) and (c), named \textit{piranha} and \textit{galvanic} etching, provide the possibility to create electrical contacts in a self-aligned way, by carrying out a metalization step prior or subsequent to etching. Here the same lithographic mask is used for the metalization and the wet etch process, so that the electrical contacts are perfectly aligned to the etched NW segment. The main difference between the first two methods is the order of execution of the metalization and etching steps, which results in significantly different chemical reaction pathways and a different etched NW geometry. The third method labeled \textit{alkaline} etching, shown in Figure~\ref{fig1}(d), exhibits a highly anisotropic etch profile, creating a conically shaped NW segment, which results in an adiabatically varying electrical potential and constriction.

\subsection{Principles of the piranha and galvanic wet etch methods}
\begin{figure}

\includegraphics{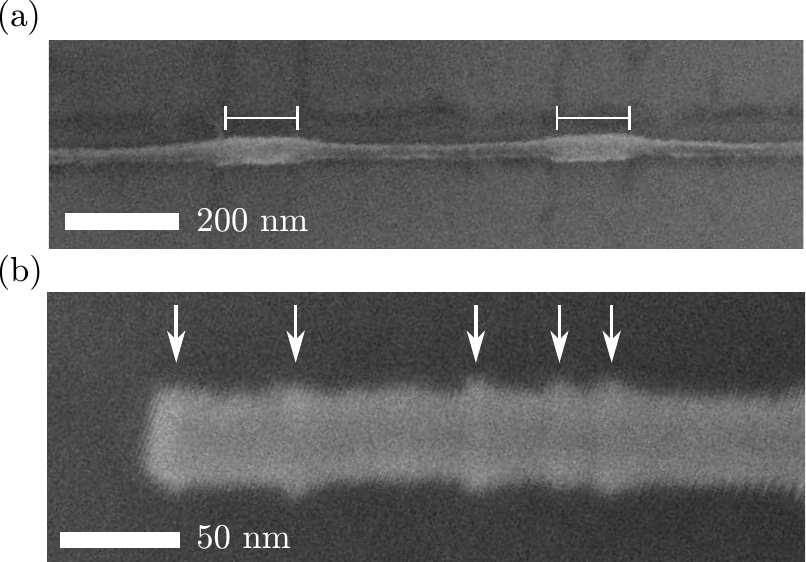}

\caption{Illustration of the basic properties of the piranha etching technique. (a) SEM image of an InAs NW undergone sulfur passivation and piranha etching resulting in an \textit{inverted} geometry. The white lines mark the NW segments in the opening of the resist mask which was used for both wet chemical processes. While the NW parts below the mask are substiantially thinned, the segments in the resist opening are practically unaffected. (b) SEM image of an InAs NW after piranha etching. Ring-like shapes are formed (marked by arrows), which we attribute to crystallographic inhomogeneities.}
\label{fig_piranha}
\end{figure}

In trial experiments we have attempted to thin certain segments of NWs defined by a lithographic mask. Narrow strip-like windows were opened in the resist over several NWs, the samples were etched in a dilute piranha solution for a few seconds (see solution I in table \ref{summary-table}), then rinsed in deionized (DI) water. After dissolving the resist with aceton the samples were inspected in an SEM. We found that the resist did not protect the nanowire from etching: starting from the opening in the resist the solution penetrated below the mask and etched the NW a few hundred nanometers below the resist.  If we carried out a sulfur passivation step (detailed below) to treat the surface of the NWs right before the piranha etching, we ended up with an \textit{inverted} geometry, shown on Figure \ref{fig_piranha}(a). In this case the NW segments in the mask opening (marked by lines) were intact, while the NW parts below the resist were still etched.

We drew two conclusions from these experiments. First, it is probable that the resist does not cover the NWs perfectly, a small gap close to the NW-wafer boundary (see Figure~\ref{fig1}(a)) allows the piranha solution to penetrate below the mask resulting in a serious underetching. Second, the NW surface treated by sulfur passivation is resistant against the piranha solution.

The surface of InAs NWs is covered by a 2-5 nm thick native oxide layer \cite{sourribes2013minimization}, the exact value depends on the details of the growth process. A water-based ammonium sulfide solution is commonly used for the removal of this layer \cite{SamuelsonPassivation}, which leaves a sulfidized surface with monosulfides, polysulfides and elemental sulfur \cite{bessolov1998chalcogenide}. The resistance of the sulfur passivated surface against the piranha solution can thus be interpreted as material selectivity, the sulfidized layer exhibits a much lower etch rate. While the piranha solution penetrates below the resist, we observe that the ammonium sulfide passivation is only effective in the resist opening. This feature can be explained with the different wetting properties of the two solutions. Furthermore, it has to be noted that while the native oxide layer is hydrophylic, the sulfidized surface is hydrophobic \cite{plis2012hydrophobic}.

Based on these findings we designed a fabrication process to achieve the desired geometry, where the NW is thinned only in the resist openings. It relies on the sulfur passivation of the whole NW surface and the reoxidization of lithographically defined segments by an oxygen plasma treatment. This process is illustrated on Figure~\ref{fig1}(b) and presented in detail in the next section.

Additionally, in some cases we have observed ring-like structures forming on the NWs after piranha etching, as shown on Figure~\ref{fig_piranha}(b), marked by arrows.  We believe that these ring-like features have a crystallographic origin. Although most of a NW has wurtzite structure, some NWs have short zinc-blende segments incorporated, which are revealed in the etching process, because they have a slower etch rate.  Having this hypothesis verified, for example by means of transmission electron microscopy, a weak piranha etching could be used as a simple alternative method for the crystallographic inspection of InAs NWs. Furthermore, the appearance of crystallographic features has implications on the chemical dissolution process, it suggests that it is reaction rate limited, not diffusion limited \cite{baca2005fabricationgaas}.

\begin{table}[]
\centering
\caption{Summary of the acidic wet etch solutions. The amount of DI water, sulfuric acid solution with 2.5 molarity and hydrogen peroxide with 30\% mass concentration is listed for each solution. Component amounts are scaled for the sake of easier comparison for 50~ml DI water.}
\label{summary-table}
\begin{tabular}{ccccc}
\hline
Solution & Type & H$_2$O & 2.5 M  H$_2$SO$_4$ & 30\% H$_2$O$_2$ \\ \hline
I & piranha  &   50 ml &  7.99 ml   &  0.59 ml           \\
II & piranha   &  50 ml      & 5 ml     & 0.5 ml              \\
III & piranha   &  50 ml      & 12.5 ml     &  1.25 ml              \\
IV & galvanic       &  50 ml      & 1.63 ml            & 0.12 ml              \\ 
V & galvanic       &  50 ml      & 0.1 ml            & -              \\ \hline
\end{tabular}
\end{table}

\subsection{Method I: Piranha etching with self-aligned contacts}
\subsubsection{Sample fabrication}

In this section we present the piranha etching technique we applied to partially thin InAs NWs. The process steps are illustrated in Figure~\ref{fig1}~(b). Having the NWs deposited on the oxidized silicon wafer, we sulfidized their surface using a highly dilute ($\sim 0.2\%$) aqueous ammonium polysulfide solution (step 1). This sulfidized layer is illustrated in orange  in the figure. 
Following reference [\citenum{SamuelsonPassivation}] we prepared first a $\approx2$~\%  solution by mixing 2~ml commercial reagent-grade 21\% (NH$_4$)$_2$S solution with 18~ml DI water and saturated it by dissolving $\approx0.19$~g elemental sulfur powder. 
The complete dissolution took for about 40~minutes at an elevated temperature. The highly dilute  (NH$_4$)$_2$S$_x$  solution  for the sulfidization was freshly prepared before every treatment by mixing 2~ml of the 2~\% solution and 8 ml DI water. The samples were immersed in this solution for 30~minutes at 40~$^\circ$C, then rinsed in DI water.
The ammonium sulfide treatment at such low concentrations is reported to be self-terminating \cite{SamuelsonPassivation}, so that the native oxide layer is completely removed yet the etching of InAs is negligible. Although the sulfidized layer is remarkably stable against reoxidation \cite{petrovykh2003passivationstab}, the samples were kept in vacuum between process steps to reduce the exposure to air.
We have observed that the position of some NWs on the substrate changes during the passivation step, some of them are even lifted from the surface. The remaining NWs are located with respect to metallic markers and a mask is created using EBL.
To make the NW sections in the mask openings susceptible for piranha etching, they were reoxidized by treating the samples in an Oxford Plasmalab Reactive Ion Etcher (step 2) \cite{Note1}. Since the parameters are machine-specific, for comparison we note that such a treatment ashes $\sim25$~nm of PMMA as measured by a profilometer. 
Following the reoxidization, the samples were etched in dilute piranha solutions (the compositions are given in rows II and III of Table~\ref{summary-table}) for times varying between 15 and 30 seconds at room temperature while stirring the solution continuously with a magnetic stirrer (step 3). Following the etching the samples were immediately rinsed in DI water.
Next, without stripping the mask, another sulfur passivation step was carried out (in this case for the sake of ohmic contacts) and metallic leads were vacuum evaporated with a 5/90/10 nm Ti/Al/Au layer structure (step 4). The evaporation is followed by lift-off (step 5).
We note that the electrical contact formation here is optional. After piranha etching the resist can be stripped off in case the leads are not needed.

\subsubsection{Results}

\begin{figure}
\includegraphics{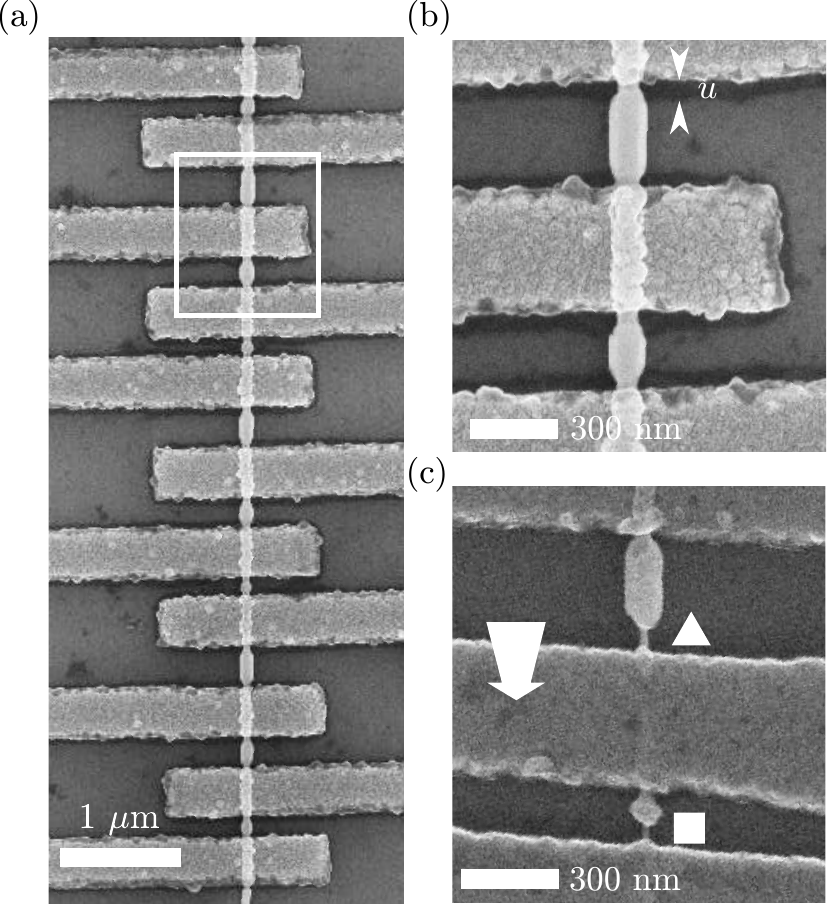}
\caption{SEM images of NWs thinned with piranha etching and equipped with self-aligned contacts. (a) NW with several contacts. (b) Zoom into the area marked by the white rectangle in (a). The length of the thinned segment reaching beyond the contact is denoted by $u$, and interpreted as the size of the undercut of the mask. (c) SEM image of another sample which was etched stronger and the metallic leads were evaporated at an angle (direction indicated by the arrow). Because of the asymmetric positioning of the electrodes the thinned segments are only revealed on the upper side of each electrode, marked by the triangle and the square.}
\label{fig_piranha_self-aligned}
\end{figure}

Following lift-off the samples were inspected in an SEM, Figure~\ref{fig_piranha_self-aligned} shows two representative samples.
In Figure~\ref{fig_piranha_self-aligned}(a) we present a NW with multiple contacts (with varying spacing), demonstrating the reliability of the method, and a zoom into the area marked by the white rectangle on Figure~\ref{fig_piranha_self-aligned}(b).
This sample was etched in solution II for 30 s at 24$^\circ$C. The original diameter of the NW measured on the intact segment is 115~nm and on the thinned part it is 70~nm.
The length $u$ of the etched segment extending from below the contact, marked by arrows is 60~nm. This length is approximately the same as the size of the undercut, and therefore we conclude that the etching is confined in the undercut.
We note that by using a thinner resist layer and/or different resist material the undercut and correspondigly, the extension of the etched segment can be made smaller, but in case a metallic contact is made in the same lithographic step, the metal--resist thickness ratio must be considered  to ensure a clear lift-off. 

The sharpness of the profile can be defined as the length of the transition along the NW axis from the thinnest to thickest (original) diameter, which is also 50--60~nm.

Figure~\ref{fig_piranha_self-aligned}(c) shows another sample which was prepared similarly as the sample in Figure~\ref{fig_piranha_self-aligned}(a-b), but was etched in a stronger etchant (see solution III in Table~\ref{summary-table}) for 20 seconds at 23$^\circ$C. Correspondingly, the etched NW segment is thinner, it has 25~nm diameter at the thinnest part.
Furthermore, since the metallic layer was evaporated at an angle (direction indicated by the arrow), the contacts are not centered on the etched segments and the thin parts are only visible on one side of them, marked by a triangle and a square.
We conclude that if a symmetric geometry is desired, as in Figure~\ref{fig_piranha_self-aligned}(a-b), the metal must be evaporated perpendicularly. On the other hand, control of the evaporation angle enables the fabrication of devices with asymmetric geometry.

\subsection{Method II: Galvanic etching with self-aligned contacts}

\begin{figure}
\includegraphics{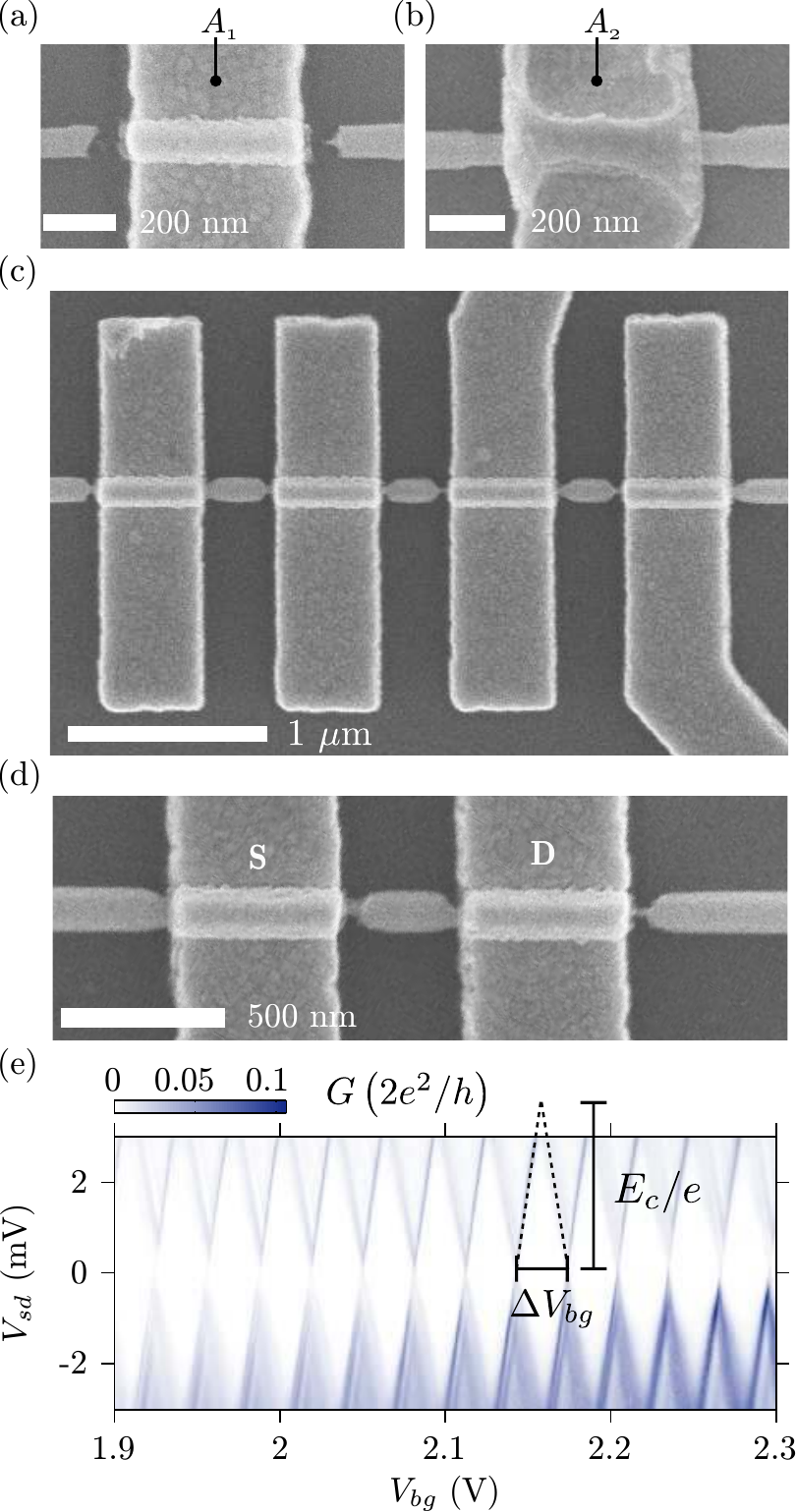}

\caption{Results of the galvanic method. (a-b) NWs with ohmic contacts after etching. Even though both NWs went through the same fabrication procedure, the etching was much stronger on the device shown in (a). We explain this finding with a local galvanic element formation which enhances the etching process, depending on the electrode area $A_i$ ($A_1 \gg A_2$). (c) NW having two electrodes with small areas on the left and two large electrodes on the right, systemic dependence of the etch rate is not observable. (d) NW device processed with the galvanic etch method, electrodes labeled as source (S) and drain (D). (e) Differential conductance of the device shown in (d) as a function of bias and backgate voltage. Dashed lines indicate the edges of a Coulomb diamond with charging energy $E_c$ and resonance spacing $\Delta V_{bg}$.}
\label{fig_galvanic_results}
\end{figure}

The piranha etching technique described in the previous section yields devices where the NW segments below the contacts are thinned, resulting in a reduced contact area between the electrode and the NW, which may be a drawback for certain applications. In this section we present a method in which this problem is resolved by switching the order of etching and metallization.   
\subsubsection{Sample fabrication}
The fabrication process is illustrated in Figure~\ref{fig1}(c). The NWs are sulfur passivated in the first step, using the same ammonium polysulfide treatment as in the previous section, then a mask is created using EBL (step 2). Unlike in the case of piranha etching, the samples were not treated with oxygen plasma after developing the mask. After development metallic contacts were vacuum evaporated with 5/120 nm Ti/Au layer structure. Without stripping the resist the NW surfaces in the undercut were reoxidized by an oxygen RIE treatment (step 3), as in the piranha etching method. Then the samples were etched in solutions IV and V described in Table~\ref{summary-table} for times ranging in 10--15 seconds, continuously stirring the solution with a magnetic stirrer (step 4). Following lift-off (step 5) the samples were inspected by SEM and electronic transport experiments were carried out at low temperatures on a subset of devices.

\subsubsection{Results}

Figures~\ref{fig_galvanic_results}(a-b) show SEM images of two different NWs which were etched in a solution prepared as given by recipe IV for 10 seconds at 23.5$^\circ$C. Although the two NWs were on the very same silicon substrate, within 500~$\mu$m distance of each other, and went through the same fabrication process, the effect of etching was dramatically different. The NW on Figure~\ref{fig_galvanic_results}(a) is completely cut on the two sides of the contact with a sharp profile, on the other hand, the NW on Figure~\ref{fig_galvanic_results}(b) is only slightly etched.
The main difference between these two NWs are the areas of the metallic electrodes, marked by $A_1$ and $A_2$ in the figure. The area $A_1$ is much larger than $A_2$ because the outer parts of electrode on \ref{fig_galvanic_results}(a), outside the SEM image, are wider and end in a large bonding pad, while the electrode on \ref{fig_galvanic_results}(b) is not connected, similarly to the two electrodes on the left side in Figure~\ref{fig_galvanic_results}(c).  While $A_2\approx 1~\mu\textrm{m}^2$, the larger electrode area $A_1\approx 10^5~\mu\textrm{m}^2$.

Furthermore, comparing the results with the previous section, although the etching solution in IV is significantly more dilute than the previous ones, it resulted in stronger etching of NWs with large electrode areas. This faster etch rate and the dependence on the electrode area has been qualitatively observed across multiple samples and we attribute it to a local galvanic element formation \cite{baca2005fabricationgaas}.

Upon close inspection of Figure~\ref{fig_galvanic_results}(a) we find that the NW segment below the contact is still present despite the complete dissolution of the neighboring etched parts.
We note that on NWs equipped with multiple electrodes differing in areas, an example shown on Figure~\ref{fig_galvanic_results}(c), we did not observe a systemic variation of the etching within the same NW as a function of the individual electrode area. We believe that as long as the NW conducts electrically, the overall electrode area is determining the etch rate, not the individual electrode areas. As the etching proceeds and the NW is cut, the NW segments decouple and the process may continue with different rates in each part. 

To reduce the etch rate and gain control over the process the etching solution was diluted and we omitted the hydrogen peroxide. Essentially, the sample shown in Figure~\ref{fig_galvanic_results}(d) was etched in a very dilute sulfuric acid solution (see solution V in Table~\ref{summary-table}) for 15 seconds at 24$^\circ$C. In the view of the principle of piranha etching, that is, the oxidization of the semiconductor surface by an oxidizing agent, and the subsequent dissolution in the acid \cite{baca2005fabricationgaas}, the etching effect in sulfuric acid alone is surprising. The complete description of the chemical processes, including the energetics and charge transfer kinetics is beyond the scope this paper, but we explain our results as follows \cite{huang2011metal,hagio1993electrode,van2001galvanic}. In the absence of hydrogen peroxide the role of the oxidizing agent is played by the solved O$_2$ and/or the H$^+$ protons originating from the dissolution of sulfuric acid. In the former case the production of OH$^-$, in the latter case the reduction of protons is taking place on the metallic electrodes (cathodic reaction), injecting holes into the semiconductor NW via the ohmic contact. The semiconductor is oxidized by the holes and dissolves (anodic reaction).

Comparing the NW on Figure~\ref{fig_galvanic_results}(c) with the one on Figure~\ref{fig_piranha_self-aligned}(a), one can observe that the galvanic method has lower reproducibility. The original diameter of the NW in Figure~\ref{fig_galvanic_results}(c) is 125~nm, while the reduced diameters vary between 85 nm and 30 nm. The rise of the variance can be attributed to the increased complexity of the underlying chemical processes.

The NW device shown on Figure~\ref{fig_galvanic_results}(d) was etched in solution V for 15 seconds at 24$^\circ$C. From left to right, the diameters on the etched parts are approximately 90~nm, 50~nm, 60~nm and 25~nm, while the original diameter is 130~nm. Although the inner two etched segments between the metallic electrodes are similar in diameter, their etch profiles are markedly different. The electric conductance of this NW was investigated using the two electrodes as source (S) and drain (D), and the doped Si substrate below 300~nm SiO$_2$ as a global backgate. The sample was cooled to 230~mK in a He-3 cryostat. The differential conductance, defined as $G=dI_{sd}/dV_{sd}$ was measured with lock-in technique, by mixing a small sine excitation $V_{ac}=10~\mu$V with a frequency 79~Hz to the dc bias and registering the ac response. The differential conductance measured as the function of dc bias voltage $V_{sd}$ and backgate voltage $V_{bg}$ is shown on Figure~\ref{fig_galvanic_results}(e). The regions of zero conductance are Coulomb diamonds, which are characteristic for the transport of quantum dots (QD). From this measurement we extract a charging energy of $E_c=3.7~\textrm{meV}$, and from the spacing of zero bias Coulomb resonances $\Delta V_{bg}=31~\textrm{mV}$, we determine the backgate capacitance $C_{bg}=e/\Delta V_{bg}=5.2~\textrm{aF}$. Assuming that the broadening of Coulomb resonances is determined by the lifetime broadening (low temperature limit), the tunnel coupling strength $\Gamma=\Gamma_s+\Gamma_d$ can be estimated \cite{beenakker1991theory}. Using the lever arm of the backgate $\alpha_{bg}=E_c/(e\Delta V_{bg})=0.12$ and the full-width-at-half-maximum of the resonance broadenings we get $\Gamma=300~\mu$eV. From excited state lines outside the Coulomb diamonds we extract an orbital level spacing of $\sim 1$~meV.

The value of the charging energy and backgate capacitance is in the typical range for InAs NWs with similar diameters and 375~nm contact spacing \cite{jespersen2006kondo}. However, the Coulomb blockade region is usually observed at negative backgate voltages. In this sample the conductance was quenched at zero backgate voltage and positive voltages were applied to induce a finite current. In general, the threshold gate voltage where the NW is depleted and the conductance in pinched was shifted to more positive values in the etched NW samples.

The Coulomb diamonds are only tilted to a small extent, indicating that the source and drain capacitances are similar.
The Breit-Wigner formula describes the shape of zero bias Coulomb resonances in the low temperature limit. In this limit the conductance maximum of a resonance is $4 \Gamma_s \Gamma_d/(\Gamma_s+\Gamma_d)^2$ in units of conductance quantum $G_0=2e^2/h$. In the symmetric case $\Gamma_s=\Gamma_d$ the fraction yields unity. The strongest zero bias resonance in the measurement has a conductance maximum $0.07~G_0$. Attributing all the reduction to the coupling asymmetry (neglecting any other resistance in series) we estimate an upper limit for $\Gamma_d/\Gamma_s\approx50$. 

All these findings are consistent with the picture that the observed quantum dot is confined by the etched NW segments next to the contact.

\subsection{Method III: Alkaline etching}
The wet etch methods presented in the previous sections allow the creation of NW geometries with sharp profiles. In this section we introduce a technique based on etching with an ammonium polysulfide solution which yields a smooth transition with conical morphology. This shape can be preferred for certain applications, for example to create quantum point contacts or nanogaps.

\begin{figure}
\includegraphics{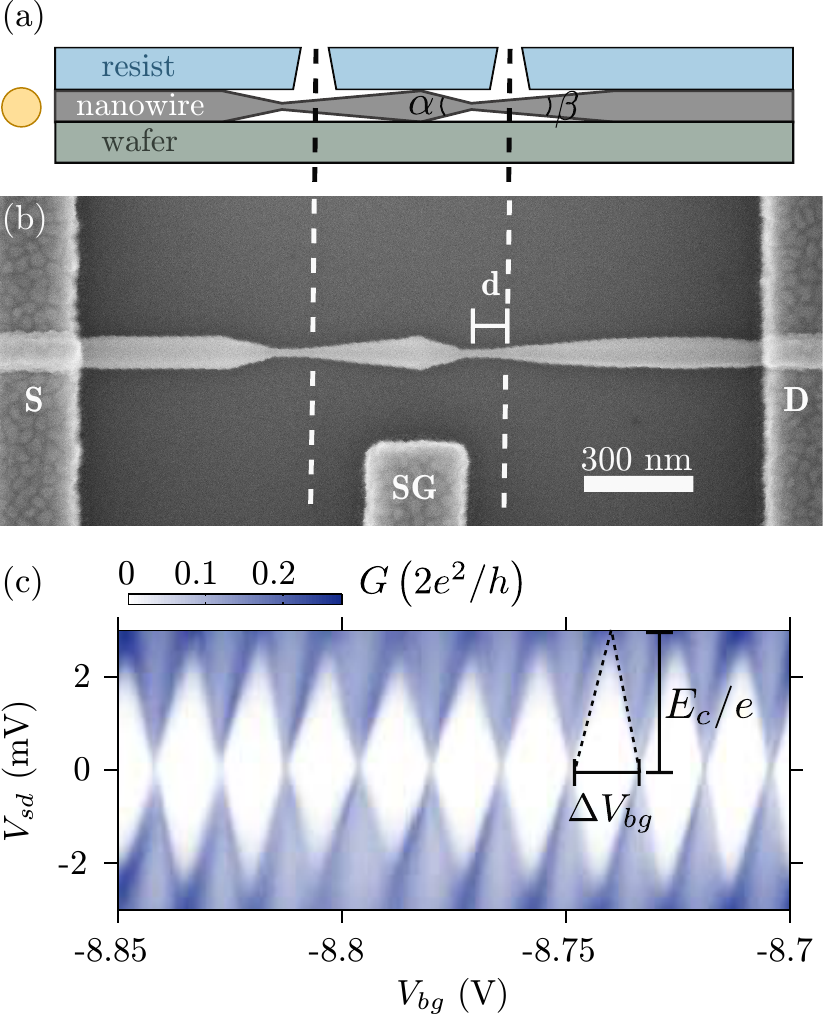}

\caption{Results of the alkaline etching method. (a) Schematic illustration of the intermediate state after the wet etch. Asymmetric and smooth etch profiles are created, $\alpha$ and $\beta$ denote the cone angles of the tapered NW segments. Yellow circle symbolizes the golden catalyst used for the NW growth, determining the crystallographic direction of the NW axis. (b) SEM image of the NW device with source (S), drain (D) electrodes and a side gate (SG). Dashed lines spanning (a) and (b) mark the position of the etch mask openings, $d\approx 110$~nm is the distance between the thinnest NW part and the corresponding etch window. (c) Differential conductance of a NW processed with the alkaline method, measured as a function of bias and backgate voltage at fixed side gate voltage $V_{sg}=0$. $E_c$ and $\Delta V_{bg}$ are the charging energy and resonance spacing, respectively.}
\label{fig5}
\end{figure}

\subsubsection{Sample fabrication}

The ammonium polysulfide solution commonly used for the removal of the native surface oxide etches the InAs material as well, although with a lower etch rate. Hence, for the patterning of the NW a solution with concentration higher than used for ohmic contact formation is more practical. 

The process is shown on Figure~\ref{fig1}(d). After transferring the NWs onto the silicon substrate the sample is spin coated with a 150~nm thick PMMA film. The resist is exposed with a 30 keV electron beam which is followed by cold development, resulting in 40~nm wide open strips crossing perpendicularly the NW. Then the sample is etched in a $4\%$ ammonium polysulfide solution saturated with sulfur. Following rinsing in DI water we strip the mask, and in a second lithographic cycle electrical contacts are made further apart from the etched locations.

\subsubsection{Results}

An SEM image of a NW fabricated with the alkaline method is shown Figure~\ref{fig5}(b). The etch profile is more smooth compared to the acidic methods, and in each etch window two conical NW segments are formed, pointing in opposite directions. At the same time the profile exhibits a remarkedly strong asymmetry, one of the cones is more elongated than the other one. This asymmetry can be characterized quantitatively by the cone angles $\alpha$ and $\beta$ shown in Figure~\ref{fig5}(a). Consistently across multiple samples, $\alpha \approx 15^\circ$ and $\beta \approx 6^\circ$ was found. In reference [\citenum{kallesoe2010integration}] a similar tapering was reported in chemical etching of GaAs segments in heterostructure NWs and was interpreted as an enhancement of an initial tapering. In our case the appearance of two segments with opposite tapering and different cone angles calls for a different explanation. We think that the lack of mirror symmetry of the wurtzite structure \cite{glasser1959symmetry} with respect to the NW axis leads to the directional dependence of the etch rate and through that to the observed asymmetric double cone morphology.
The diameter of the narrowest part of the NW is 20~nm, while the original diameter is 90~nm. Counterintuitively, the constriction is made at a distance $d\approx 110$~nm from the center of the etch window (see Figures~\ref{fig5}(a-b)). The etch rate calculated using the reduction of diameter at this thinnest part is $\approx$5~nm/min.

Conductance measurements have been carried out to investigate the transport properties of a NW device with such a geometry. The differential conductance measured at $T=230$~mK in the Coulomb blockade regime is shown on Figure~\ref{fig5}(c). Repeating the same analysis as in the previous section, we get $E_c=3$~meV, $\Delta V_{bg}=15$~mV, $\alpha_{bg}=0.2$, $C_{bg}=e/\Delta V_{bg}=11$~aF and $\Gamma=170~\mu\textrm{eV}$. Additionally, the ratio of the backgate and side gate lever arm  is determined from the conductance measured at zero bias as the function of $V_{bg}$ and $V_{sg}$ (not shown), yielding $\alpha_{bg}/\alpha_{sg}=3.5$.
Compared to the device in Figure~\ref{fig_galvanic_results}(c), the backgate lever arm and capacitance is increased. This is the consequence of the reduced screening of the leads placed further from the QD.
Based on the same symmetry considerations of the Coulomb diamonds as in the previous section, we conclude that the QD is formed between the two constrictions. This is consistent with the lever arm ratio, both gates are at a distance $\sim$300~nm from the QD, but the dielectric permittivity $\sim3.9$ of the silicon dioxide increases the effect of the backgate.

The highest zero bias conductance maximum was {$\sim0.15~G_0$}, and the corresponding estimate of coupling strength asymmetry is $\Gamma_s/\Gamma_d\approx25$. Since the contacting electrodes are far apart from the barriers defining the QD, it is plausible to assume that the NW segments between the QD and the leads contribute to the resistance substantially, making the asymmetry estimation pessimistic.

\section{Conclusions}
We have shown three wet chemical etch methods for the post-growth engineering of InAs NWs. These methods allow the thinning of NWs on lithographically defined segments. Two of them are well-suited to fabricate electrical contacts on top or next to the etched parts in a self-aligned way, the third one creates a smooth conical NW profile.

The piranha and alkaline methods are found to be robust and reliable. In these methods the control of the etched NW diameter is limited by the deviation of diameter in the growth. The individual NW diameter can be measured by an SEM prior to etching and the etching time can be adjusted. The galvanic etching proved to be less reproducible due to the complexity of the chemical processes making the etch rate depend on the contact electrode surfaces. The reproducibility in this case could be improved by breaking the sample fabrication presented in Figure~\ref{fig1}(c) into two lithographic steps, first creating electrodes with a small, well-defined contact area for the etching, and in a second step adding large electrodes necessary for wire bonding. We note that this modification does not sacrifice the self-aligned nature of the etching.

Low temperature experiments show that NWs with etched segments have stable transport characteristics and can serve as building blocks of quantum electronic devices. As an example we show that QDs can be formed between two etched segments. The demonstrated post-growth wet etch methods could be applied in other quantum devices, for example forming QPCs with adiabatic confinement, strongly motivated by the search for Majorana Fermions in such wires \cite{wimmer2011quantum}.

\section{Acknowledgments}
We thank  Péter Makk, Bálint Fülöp, Gábor Fábián, Zoltán Scherübl, Thomas Sand Jespersen, Rawa Tante and Morten H. Madsen for useful discussions and experimental assistance.

We gratefully acknowledge the financial support by the EU FP7 project SE$^2$ND, the EU ERC projects CooPairEnt and QUEST,
the Swiss NCCR Quantum, the Swiss SNF, the Danish Research Councils, the Danish National Research Foundation, and the Hungarian
Grant No. OTKA K112918. S.~C. was supported by the Bolyai Scholarship and G.~F. was a SCIEX fellow (project NoCoNano).

\end{document}